# Equity Scores for Public Transit Lines from Open-Data and Accessibility Measures

**Amirhesam Badeanlou**
Department of Environment, Land and Infrastructure Engineering
Polytechnic University of Turin, Turin, Italy, 10129
Email: amirhesam.badeanlou@studenti.polito.it

**Andrea Araldo**
Télécom SudParis
Institut Polytechnique de Paris, France, 91120
Email: andrea.araldo@telecom-sudparis.eu

**Marco Diana**
Department of Environment, Land and Infrastructure Engineering
Polytechnic University of Turin, Turin, Italy, 10129
Email: marco.diana@polito.it

**Vincent Gauthier**
Télécom SudParis
Institut Polytechnique de Paris, France, 91120
Email: vincent.gauthier@telecom-sudparis.eu

Word Count: 2465 words + 1 table (250 words per table) = 2,715 words

*Submitted [Submission Date]*

A.Badeanlou, A.Araldo, M.Diana, and V.Gauthier

**ABSTRACT**

**Current transit suffers from an evident inequity: the level of service of transit in suburbs is much less satisfying than in city centers. As a consequence, private cars are still the dominant transportation mode for suburban people, which results in congestion and pollution. To achieve sustainability goals and reduce car-dependency, transit should be (re)designed around equity. To this aim, it is necessary to (i) quantify the "level of equity" of the transit system and (ii) provide an indicator that scores the transit lines that contribute the most to keep transit equitable. This indicator could suggest on which lines the transit operator must invest to increase the service level (frequency or coverage) in order to reduce inequity in the system.**

**To the best of our knowledge, this paper is the first to tackle (ii). To this aim, we propose efficient scoring methods that rely solely on open data, which allows us to perform the analysis on multiple cities (7 in this paper). Our method can be used to guide large-scale iterative optimization algorithms to improve accessibility equity.**

A.Badeanlou, A.Araldo, M.Diana, and V.Gauthier

**INTRODUCTION**

Private cars are among the main sources of pollution (1) (61% of transportation emissions in Europe (2) ) and they are still the dominant transportation mode in many parts of metropolitan areas, in particular the suburbs (12). The main reason is that in such areas travelers do not often have a competitive alternative to their private cars for commuting, as the level of service offered by Public Transit (PT) is often poor therein, i.e., frequency and coverage are insufficient, resulting in high waiting and walk-to-station times. All policies devoted to decreasing the use of cars, e.g., higher car ownership taxes, tolls, fuel prices, are destined to fail if PT service remains poor in suburbs, as travelers will anycase be obliged to use their private cars, due to lack of convenient alternatives. Therefore, to achieve car-independent sustainable mobility, it is urgent to reduce the inequity of the geographical distribution of PT level of service.

PT level of service can be measured based on accessibility indicators (5,7). Several indicators have been proposed in the literature (4, 5, 10). In broad terms, the accessibility of a location measures how well it is connected to the rest of its surrounding region. By observing the geographical distribution of accessibility over city centers and the suburbs, we can uncover public transit service inequity.

So far transit has not been designed with accessibility equity in mind: Transit Network Design Problems (TNDP) usually decide line frequencies and routes with the aim to optimize some generalized cost function, which includes the sum of all operation, capital and user costs (9,11). For the aforementioned reasons, we believe instead that newer methods must be proposed where accessibility equity is the main optimization objective. This paper is the first step toward such equity-maximizing transit design methods. Indeed, in order to construct such methods, we first need to understand what are the most relevant public transport lines that help preserve equity.

To this aim, we propose in this paper a method to compute per-line *equity scores*. This information may guide the strategic decisions of transit operators, to choose which lines should be primarily enhanced (e.g., increasing their frequency and connectivity). We first compute equity score as the difference in the Gini inequity index when adding/removing a certain line. Then, we propose an approximation of such a score that is two orders of magnitude faster. The computation time-efficiency of our method enables its use in optimization loops for transit network design problems (9), where thousands of alternative transit configurations may be compared, thus requiring the computation of performance indicators to be extremely fast.

Much previous work on accessibility requires rich datasets, e.g., household income and employment, business location and type and surveys. Such data are usually not publicly available or difficult to use (they may be in a non-standard format or in local languages). Consequently, previous work usually just focuses on one scenario (8, 13). We choose instead to solely rely on open data in standard form, which, as in (5,7), allows us to easily repeat our analysis on multiple cities. We showcase our methodology in Manchester, Turin, Aachen, Vienna, Helsinki, Berlin and Budapest. Our code is available at (6).

**PRELIMINARY DEFINITIONS**

**Accessibility**

We define the accessibility of a certain place as a measure of how easy it is for a passenger to travel and reach opportunities (places, people, businesses), starting from that place, by using transit. A well-connected place, close to a station where several lines pass, with high service frequency, will obviously enjoy high accessibility.

Among the many accessibility scores defined in the literature (4,5,10), we choose the *sociality score* presented in (5) as (i) it can be computed solely based on open data and (ii) it takes into account social interaction. We here broadly explain its calculation, skipping the detailed formalization. We partition the area under study with a hexagonal tessellation, with hexagons $\lambda \in \Lambda$ of 1Km per side. We associate to each hexagon the population resident therein, using open data from Eurostat[1]. We remove the hexagons whose population density is not significant (less than 100 residents per $Km^2$). For each hexagon $\lambda$, we consider all the possible trips departing at time $t$=8AM from its barycenter and lasting at most $T$=1$h$. Let us denote with $\mathcal{L}$ the set of transit lines.
The calculation of trips within public transit is based on General Transit Feed Specification (GTFS), a standard data format to describe the detailed transit schedule. Many transit operators now publish their schedules in such a format. The GTFS data of a certain urban area consists of several files, among which `stops.txt` shows the exact location of each transit stop (either bus stop or metro/train station) and `stop_times.txt` shows which line is serving each stop and at which time, which allows to reconstruct the trajectory of each transit vehicle. As for the calculation of the time to walk to and from stations, we resort to OpenStreetMap and Open Source Routing Machine (OSRM)[2].

Thanks to this information, starting from the barycenter of any hexagon $\lambda \in \Lambda$, we can compute what are all the hexagons that can be reached within time range $T$. A curve (isochrone) is then calculated to include the reached hexagons. The accessibility score $a(\lambda, t, \mathcal{L})$ is the number of individuals living inside the isochrone. In other words, the accessibility score indicates how many people can be reached within time $T$ (in this sense it takes into account social interaction).

Figure 1 represents the accessibility scores we computed in Paris region. It is evident that hexagons in the city center enjoy much better accessibility than the first and the second ring around Paris. It is interesting to notice a certain correlation with the modal share in Paris (14): cars dominate where accessibility is low.

---

[1] https://ec.europa.eu/eurostat/web/gisco/geodata/reference-data/population-distribution-demography/geostat
[2] https://github.com/Project-OSRM/osrm-backend

A.Badeanlou, A.Araldo, M.Diana, and V.Gauthier

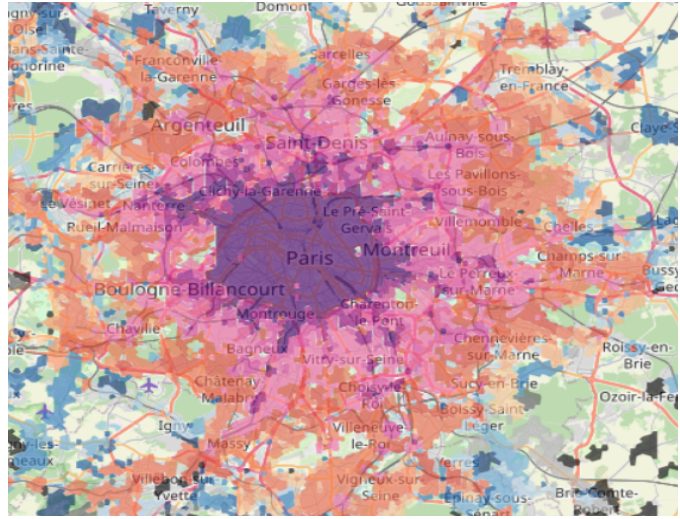

**Figure 1** Accessibility score in Paris (source: (7))

Observe that if we increase the set of available transit lines, the set of reachable places reachable from any hexagon within *T* increases (or at least it does not decrease) and thus the isochrone increases as well (or at least it does not decrease). Therefore, the following proposition, which will be useful later, holds.

***Proposition 1*** *Fixing any λ and t, $a(\lambda, t, \mathcal{L})$ is an increasing set function, i.e., $a(\lambda, t, \mathcal{L}') \leq a(\lambda, t, \mathcal{L})$ if $\mathcal{L}' \subset \mathcal{L}$.*

**Inequity**

Having defined accessibility, we now want to quantify the inequity in its distribution. A myriad of metrics to quantify inequity exist (15), among which we are interested in *spatial inequity* definitions, as (i) it is relevant for transportation studies and (ii) does not need in-depth socio-demographic data that are difficult to obtain. In particular, as in our preliminary work (7) in which the reader can find all the details of calculation, we quantify inequity as the Gini index computed based on the the Lorenz curve. The Lorenz curve is an economic tool used to quantify how a certain resource is distributed among the population. Our idea is to consider accessibility as the resource to be shared. To construct the Lorenz curve, we assume that each individual *p* living in the hexagon λ benefits from the accessibility of that hexagon, i.e., denoting with $p \in \lambda$ any individual living in the hexagon λ, we set $a(p) = a(\lambda)$. We denote with $\mathcal{P}$ the set of all individuals. We then sort all the individuals $p_1, \cdots, p_{|\mathcal{P}|}$, from the worst to the best accessibility, such that $a(p_i) \leq a(p_{i+1})$. We put in the x-axis the sorted individuals and for each individual $p_i$ the corresponding value in the Lorenz curve is

$$L(p_i) = \frac{1}{K'} \cdot \sum_{j=1}^{i} a(p_j),$$

where constant *K'* is a normalization factor, so that the Lorenz curve goes from *0* to *1*, i.e.,

$$K' = \sum_{j=1}^{|\mathcal{P}|} a(p_j).$$

We also normalize the x-axis, so that it goes from *0* to *1*.

The Lorenz curve for Paris is depicted in (**Figure 2**). Note that 80% of individuals only get 40% of the accessibility, while only 20% of individuals (range 0.8 - 1 in the x-axis) enjoy all the rest of accessibility. In the "perfect equity" case, instead, we should have that $x$% of individuals should get $x$% of accessibility, and the Lorenz curve would be the straight 45° orange curve. The level of inequity can be thus quantified as the area between the perfect equity curve and the actual Lorenz curve. We divide such an area by 0.5 (which is the area under the perfect equity curve), in order to get a number in [0,1]. Such a number is called the *Gini index*.

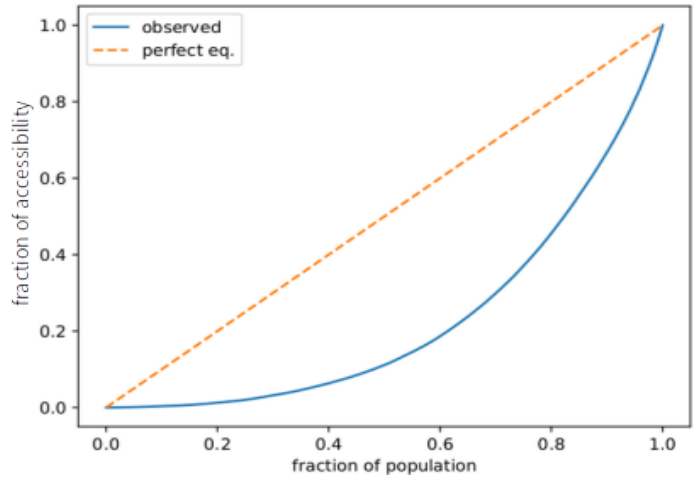

**Figure 2 Lorenz curve for Paris (from (7))**

**EQUITY SCORE**
Let us consider the set of lines $l \in \mathcal{L}$, either bus, metro, or commuter rail. We want to associate an *equity score* to any line $l$, to measure how important its contribution is to the overall transit equity.
To this aim, we compute the contribution of line $l$ to the Gini index. Due to Prop.1, passing from $\mathcal{L}\setminus\{l\}$ to $\mathcal{L}$ surely improves the accessibility of all hexagons. However, it is important to understand how such an improvement is distributed across hexagons, i.e., for which hexagons this improvement is very high and for which others it is instead negligible.

Suppose that, when passing from $\mathcal{L}\setminus\{l\}$ to $\mathcal{L}$, the highest accessibility improvement is observed for the "unfortunate" hexagons, i.e., the ones suffering from low accessibility, whose population is on the left of the Lorenz curve (**Figure 2**). This would suggest that $l$ has a beneficial impact on equity. Therefore, in this case, the left part of the Lorenz curve would "inflate" while, due to its scaling to 1, the right part would deflate. As geometrical evidence, the Lorenz curve would approach the perfect equity curve and the Gini index would decrease. In this case, we can say that $l$ positively contributes to equity.

If instead, when passing from $\mathcal{L}\setminus\{l\}$ to $\mathcal{L}$, the highest improvement is observed in the "fortunate" hexagons, i.e., the ones enjoying high accessibility, then the opposite considerations would hold: in this case line $l$ would not be beneficial to equity and the Gini index would increase. Note that, in this case, it would not mean that line $l$ is useless or deleterious. It would just mean that it does not help improve equity (although it may improve average accessibility or overall travel time). This would not mean that $l$ is to be eliminated but that, if a transit operator has a limited budget to improve its offer, it should prefer to invest in improving other lines rather than $l$, e.g., by increasing frequency or introducing advanced technology like automation.

A.Badeanlou, A.Araldo, M.Diana, and V.Gauthier

We have thus shown that the change in the Gini index when passing from $\mathcal{L}\setminus\{l\}$ to $\mathcal{L}$ gives an indication of its contribution to equity: if the Gini index increases a lot, we can infer that $l$ worsens inequity, otherwise, $l$ positively contributes to equity. More formally, let us denote with $G(\mathcal{L}, t)$ the Gini index when all lines are active, considering departures at time $t$. Let us denote with $G(\mathcal{L}\setminus\{l\}, t)$ the Gini index without line $l$. We define the *equity score* as

$$\Delta G(l, t) = G(\mathcal{L}\setminus\{l\}, t) - G(\mathcal{L}, t) \qquad (1)$$

If $\Delta G(l, t)$ is largely positive, then $l$ is important for equity at time $t$, while if $\Delta G(l, t)$ is small or negative, that line is irrelevant in terms of equity.

**Efficient calculation of equity scores**

The equity score, as defined before, might be difficult to compute, as it requires a relatively large amount of computation, which may be prohibitive in large cities, as we will see in the numerical results. To compute the equity score $\Delta G(l)$ of any line $l$, we need to

1. Compute the accessibility score $\{a(\lambda, t, \mathcal{L}) | \lambda \in \Lambda\}$ of all hexagons. We do so based on (5), which in turn is based on the Connection Scan algorithm (16).

2. Construct the Lorenz curve, as discussed in the previous section.

3. Compute the Gini index $G(\mathcal{L}, t)$.

These three steps may take several hours in big cities. Then, we would need to repeat the previous 3 steps over and over, removing every time a line $l \in \mathcal{L}$, so as to compute $G(\mathcal{L}\setminus\{l\}, t)$ and thus $\Delta G(l, t)$. Doing so in big cities with thousands of lines may require tens of days (Table 3).

This computational complexity is impractical if one wants to employ equity scores within an optimization loop, in which lines are iteratively modified in order to get, at the end of the loop, a transit structure with better equity. We thus propose here a method to compute an alternative score $e(l, t)$, $\forall l \in L$, more computationally efficient than $\Delta G(l, t)$, while preserving its meaning, i.e., indicating the contribution of each line toward equity.

In particular, we instrument the earliest arrival path algorithm with an *importance* function $i(\lambda, l, t)$ which we initialize to $i(\lambda, l, t) = 0, \forall \lambda \in \Lambda, l \in \mathcal{L}$. While constructing the earliest arrival time paths from $\lambda$ to any other hexagons, departing at time $t$ and within travel time T, we increment $i(\lambda, l, t)$ by the distance traveled via line $l$ (0 in case $l$ is never used). In this way, we give more weight to the lines that take a user departing from hexagon $\lambda$ as furthest as possible. Indeed, such lines are likely to be those that contribute the most to the accessibility of $\lambda$. Therefore, at the end of the earliest arrival path computation, we obtain, with almost-zero additional computational cost, the function $i(\lambda, l, t)$ telling how important the contribution of each line $l$ is for the accessibility of any hexagon $\lambda$, considering departure time $t$.

We now order hexagons $\lambda = 1, \cdots, |\Lambda|$ from the "worst" (the one with the lowest accessibility) to the "best" (the one with the highest accessibility). We define the *cumulative importance function* as

$$I(\lambda, l, t) = \sum_{\lambda'=1}^{\lambda} i(\lambda', l, t).$$

It expresses the importance of line $l$ for the $\lambda$ worst hexagons, which are $\lambda' = 1, \cdots, \lambda$, following the order above. We depict an example in (**Figure 3**). Note that the absolute values of cumulative importance do not have any physical meaning. We use them just to compare their values so as to score lines.

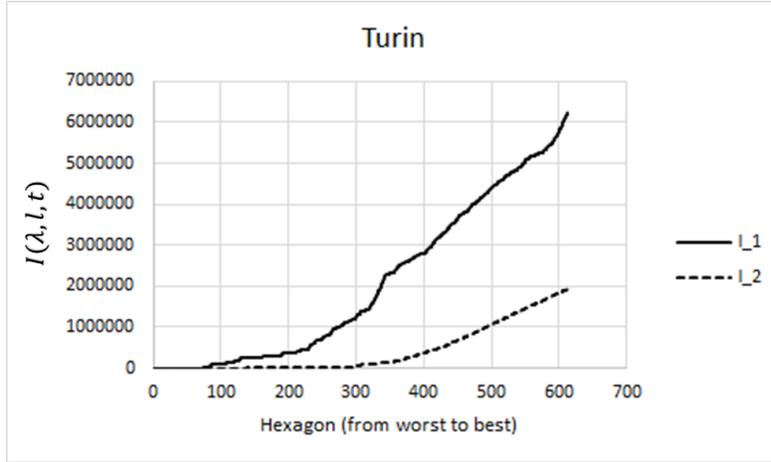

**Figure 3 Cumulative importance function for two lines $I(\lambda, l_1, t)$ and $I(\lambda, l_2, t)$ as a function of $\lambda$. Line $l_2$ contributes to equity less than $l_2$**

It is evident that a line whose cumulative importance is very high for the worst hexagons (e.g., $l_1$ of **Figure 3**) is beneficial for equity, as it allows people in disadvantaged hexagons to go relatively far. On the contrary, a line whose cumulative importance function remains low for the worst hexagons and only increases for the best hexagons does not contribute to equity. For instance, in (**Figure 3**), line $l_2$ does not contribute at all to the accessibility of the worst 300 hexagons.

We take $\lambda^{65th}$, i.e., the 65th percentile hexagon, and we define the *equity score* as

$$e(l, t) = I(\lambda^{65th}, l, t) = \sum_{\lambda'=1}^{\lambda^{65th}} i(\lambda', l, t).$$

Observe that $e(l, t)$ is an alternative equity score to $\Delta G(l, t)$ to compute. Indeed, instead of repeating a full accessibility computation (the 3 steps explained above) $|\mathcal{L}|$ times (one per line), we just need to compute accessibility once and, *while* doing so, we also opportunistically compute set $\{e(l,t)|l\in\mathcal{L}\}$. Therefore, the computation of all $e(l, t)$ scores is $|\mathcal{L}|$ times faster than $\Delta G(l, t)$.

However, $\Delta G(l, t)$ directly accounts for the variation of Gini inequity index induced by $l$, while $e(l, t)$ is constructed heuristically and may appear an arbitrary measure. The corroborate such measure, we thus need to verify whether there is a correlation between $e(l, t)$ and the change in the Gini inequity index. The next section will thus be dedicated to show the Claim below.

*Claim 2* The equity score $e(l)$ basically captures the same information of $\Delta G(l)$, in a much more computationally efficient way.

A.Badeanlou, A.Araldo, M.Diana, and V.Gauthier

As for the choice of the 65th percentile, we empirically observed that it verifies the claim above better than other values.

**EVALUATION**

Our method to compute the equity score is solely based on open data and can thus be easily replicable in different cities. To showcase its generality, we focus on 7 European cities of different characteristics and transit network sizes. They are listed in (**Table 1**).

**Table 1 Considered cities.**

| City | Num. of lines | Num. of hexagons | Gini Index |
|---|---|---|---|
| Manchester | 637 | 2642 | 0.445 |
| Turin | 207 | 926 | 0.374 |
| Aachen | 354 | 2547 | 0.462 |
| Vienna | 259 | 596 | 0.214 |
| Helsinki | 469 | 2146 | 0.305 |
| Berlin | 1187 | 3038 | 0.375 |
| Budapest | 281 | 1003 | 0.352 |
| Paris | 1890 | 11640 | 0.514 |

In the following computation, we consider departure time $t = 8AM$. Observe that by repeating the same analysis with different departure times we get different numbers, as transit line frequencies change over the day. However, the findings related to the methodology do not change.

**Geographical dependence of accessibility**

As preliminary observation, we report there is a strong dependence between the accessibility of a hexagon and its position in the cities under study (examples are in **Figure 4**). The hexagons suffering from lower accessibility are the ones further from the city center.

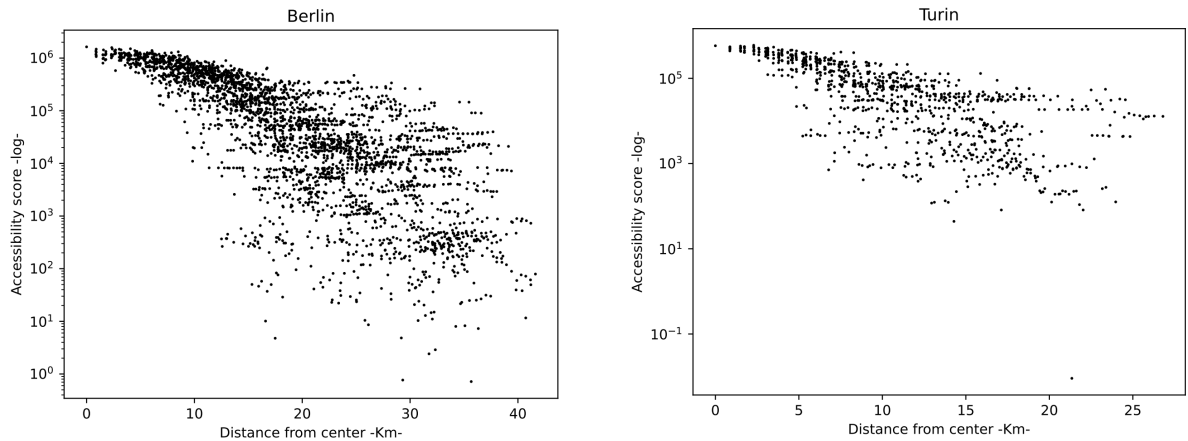

**Figure 4 Dependence between accessibility scores and distance from the center.**

**Correlation between equity scores**

We now empirically confirm that Claim 2 holds. The correlation between the two equity scores $\Delta G(l,t)$ and $e(l,t)$ appears in the scatterplots of (**Figure 5**) for two exemplary cities, where each point corresponds to a line $l$, whose x coordinate is $\Delta G(l,t)$ and y coordinate is $e(l,t)$.

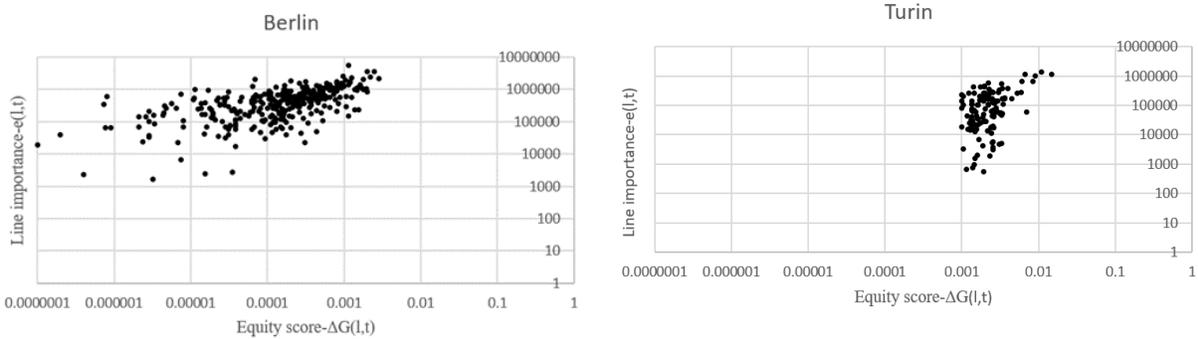

**Figure 5 Equity scores**

To generalize this observation, (**Table 2**) lists the Pearson's correlation coefficient between the values $\Delta G(l,t)$ and $e(l,t)$, for all lines $l \in L$. In all 7 considered cities, we observe a relatively strong correlation. The fact that the two scores carry very similar information is confirmed by the extremely low *p*-values.[3]

**Table 2 Correlation between $e(l,t)$ and $\Delta G(l,t)$**

| City | Pearson Coeff | *p*-value |
|---|---|---|
| Manchester | 0.62 | 6E-18 |
| Turin | 0.34 | 2E-11 |
| Aachen | 0.47 | 6E-45 |
| Vienna | 0.58 | 1E-22 |
| Helsinki | 0.43 | 1E-12 |
| Berlin | 0.42 | 1E-20 |
| Budapest | 0.66 | 3E-112 |

A concrete example in Turin and Berlin is given in (**Figure 6**), where we have on the left lines contributing positively to equity and on the right lines not contributing to equity. Note that both $\Delta G(l,t)$ and $e(l,t)$ are higher for the lines on the left than on the right.

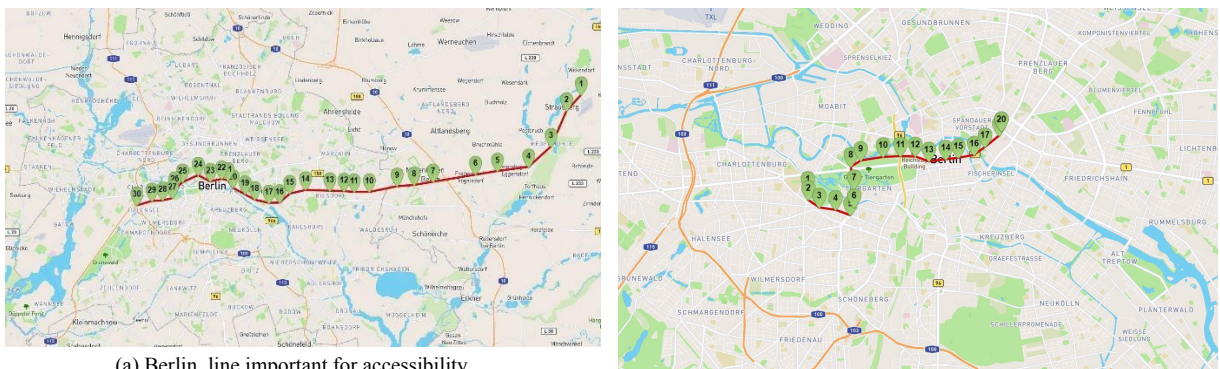

(a) Berlin, line important for accessibility.

---

[3] Recall that the p-value can be seen as the probability of observing the measured correlation "by chance", under the hypothesis that there is actually no dependency between *e(l,t)* and *ΔG(l,t)*. Such probability is in our case almost zero, which means that the correlation between *e(l,t)* and *ΔG(l,t)* is statistically very significant.

A.Badeanlou, A.Araldo, M.Diana, and V.Gauthier

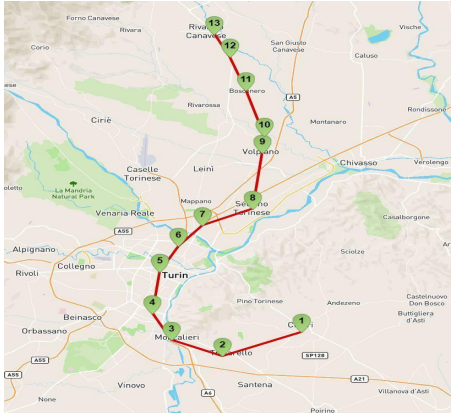

(S-Bahn Berlin GmbH - Rail Line S5)
$\Delta G(l, t) = 7.6 \cdot 10^{-4}$, $e(l, t) = 8.0 \cdot 10^{6}$

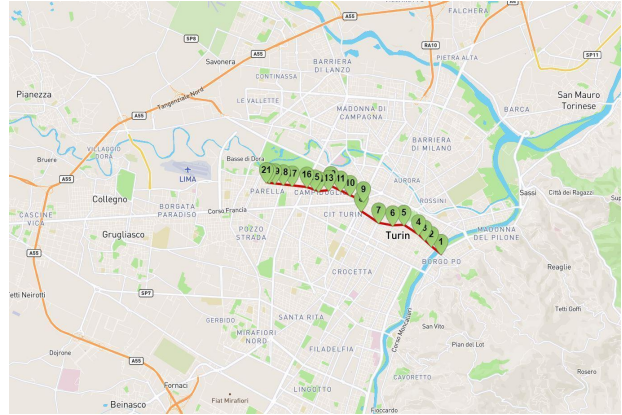

(b) Berlin, line not contributing to accessibility.
(Berliner Verkehrsbetriebe - Bus Line 100)
$\Delta G(l, t) = 7.3 \cdot 10^{-7}$, $e(l, t) = 2.0 \cdot 10^{6}$

(c) Turin, line important for accessibility.
(Train SFM1)
$\Delta G(l, t) = 1.3 \cdot 10^{-3}$, $e(l, t) = 2.8 \cdot 10^{6}$

(d) Turin, line not contributing to accessibility.
( Tram line 13 )
$\Delta G(l, t) = -2.2 \cdot 10^{-4}$, $e(l, t) = 3.6 \cdot 10^{5}$

**Figure 6 Example of equity scores**

Note that, for both $\Delta G(l, t)$ and $e(l, t)$ scores, their absolute numerical value does not matter and has no physical meaning. The scores are only interesting to understand which lines are more important than others in terms of equity: the relative ranking is interesting, rather than the absolute value.

**Computational gain**

The reason for using $e(l, t)$ instead of $\Delta G(l, t)$ is because the former is $|\mathcal{L}|$ times faster to compute. We show the computation gain of using $e(l, t)$ instead of $\Delta G(l, t)$ in (**Figure 7**), calculated as the time to compute the latter over the time to compute the former.

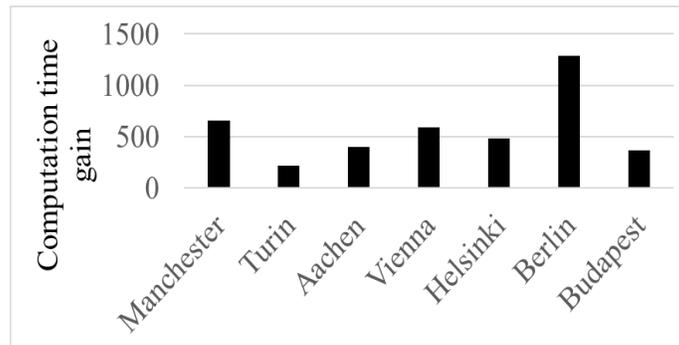

**Figure 7 Computation time gain**

Computing $e(l, t)$ is several hundred times faster than $\Delta G(l, t)$. To give an idea of the order of magnitude of computation time, we report in (**Table 3**) the time needed to compute the Gini index $G(\mathcal{L}, t)$, the time needed to compute the set of all equity scores $\Delta G(l, t)$ and to compute the set of all equity scores $e(l, t)$. The results are obtained on a virtual machine using 32 AMD EPYC 7532 CPUs and 64GB memory. Note that the time to compute $\{\Delta G(l, t)|l \in \mathcal{L}\}$ is estimated (multiplying by $|\mathcal{L}|$ the time for computing $G(\mathcal{L}, t)$ ). It is evident that for bigger cities like Paris computing $\{\Delta G(l, t)|l \in \mathcal{L}\}$ would be impractical. In smaller cities, one should wait some hours or days to get $\{\Delta G(l, t)|l \in \mathcal{L}\}$, which might be acceptable in some cases. However, it would be impossible to tolerate such high times if one wants to use the information from equity scores within some optimization loops for network design. If we suppose to use metaheuristics, e.g., genetic algorithms, or artificial intelligence, e.g., reinforcement learning, for network optimization, they would require hundreds or thousands iterations. In this case, it would be impossible to compute $\{\Delta G(l,t)|l\in\mathcal{L}\}$ at every iteration and it would be better to resort to $\{e(l,t)|l\in\mathcal{L}\}$, which is several orders of magnitude faster to compute.

**Table 3 Computation time.**

|  | $G(t)$ | $\{\Delta G(l,t)|l\in\mathcal{L}\}$ | $\{e(l,t)|l\in\mathcal{L}\}$ |
|---|---|---|---|
| Manchester | 300 sec | 2.2 days | ~5 min |
| Turin | 60 sec | 3.5 hours | ~1 min |
| Aachen | 90 sec | 8-9 hours | ~90 sec |
| Vienna | 40 sec | 2.9 hours | ~40 sec |
| Helsinki | 95 sec | 12.3 hours | ~ 95 sec |
| Berlin | 210 sec | 2.8 days | ~4 min |
| Budapest | 50 sec | 3.9 hours | ~ 50 sec |
| Paris | 720 sec | 15 days | ~12 min |

**CONCLUSIONS**

We proposed a methodology to compute an *equity score* representing the contribution to accessibility equity provided by each line. This may guide planners in choosing lines on which to concentrate investment, with the aim to improve spatial equity in the geographical distribution of public transit accessibility. This may be particularly relevant in budget constrained scenarios. Indeed, if the budget were infinite, achieving a good accessibility distribution would be easy: massive investment could be dedicated to increasing frequencies (and thus the fleet) on all lines. However, due to the limitedness of budget, operators need to make choices and prioritize certain lines over others. We advocate that such prioritization choices should not be (solely) made to improve average or total accessibility or social welfare. We believe instead that it is fairer to prioritize those lines that favor accessibility.

In our future work, we will also consider how to enrich accessibility computation with additional (non-public) data, like employment, business locations and types, when available.

This work is a first step toward methodologies to construct future generation public transit, where accessibility equity is the main criterion to optimize. We will consider in the future the combination of fixed lines and flexible modes (3,17, 18) (mobility on demand or demand-responsive buses) towards this aim.

**CONTRIBUTIONS OF THE AUTHORS**

All authors contributed to the general concept and reviewed the paper. A.A. proposed the basic idea of equity scores. A.B. contributed to the definition of the equity scores. He did all the work related to implementation and analysis. M.D. and V.G. provided scientific guidance and supervision.

**ACKNOWLEDGEMENT**

This work has been supported by The French ANR research project MuTAS (ANR-21-CE22-0025-01) and partially performed on the THD Platform of Télécom SudParis.